\documentclass[12pt,preprint]{aastex}

\shorttitle{Lithium in an AGB Star in M3}
\shortauthors{Givens \& Pilachowski}

\begin{document}

\title{THE ABUNDANCE OF LITHIUM IN AN AGB STAR IN THE GLOBULAR CLUSTER M3
(NGC 5272)}

\author{R. A. Givens \& C. A. Pilachowski\altaffilmark{1}}
\affil{Astronomy Department, Indiana University Bloomington,
    Swain West 319, 727 East Third Street, Bloomington, IN 47405-7105, 
USA; cpilacho@indiana.edu, rashagd12@yahoo.com}

\altaffiltext{1}{Based on observations obtained with the Apache Point
Observatory 3.5 meter telescope, which is owned and operated by the 
Astrophysical Research Consortium.}

\begin{abstract}
A survey of red giants in the globular cluster M3 with the Hydra
multi-object spectrograph on the WIYN 3.5-m telescope indicated a 
prominent Li~I 6707 $\AA$ feature in the red giant vZ 1050. 
Followup spectroscopy with the ARC 3.5-m telescope confirmed this
observation and yielded a derived abundance of 
A(Li)$_{NLTE}$~=~1.6~$\pm$~0.05. 
In addition, the high oxygen and low sodium abundances measured from the same
spectrum suggest that vZ~1050 is a first generation cluster star.
The location vZ~1050 above the horizontal branch and blueward
of the red giant branch in the cluster's color-magnitude diagram
places vZ~1050 on M3's asymptotic giant branch.  
The likely source for the enhanced lithium abundance
is the Cameron-Fowler mechanism operating in
vZ~1050 itself.

\end{abstract}


{\it Facility:} \facility{WIYN: 3.5m (Hydra), ARC: 3.5m (ARCES)}

\newpage

\section{Introduction}

Since the discovery of the first lithium-rich red giants in globular clusters (NGC 362: Smith et al. 1999, M3: Kraft et al. 1999),
similar lithium rich giants have been identified in other
clusters, including a second lithium-rich red giant in 
NGC 362 (D'Orazi et al. 2015) and one in NGC 4590 (Ruchti et al. 2011).
Studies of lithium in main sequence turnoff stars in globular clusters
have also found Li-rich stars even at this earlier evolutionary phase
(M92: Boesgaard et al. 1998, NGC 6397: Koch et al. 2011; M4:
Monaco et al. 2012).  
Koch et al. discovered a turnoff star in the metal-poor globular cluster
NGC 6397 with a lithium abundance A(LI)$_{NLTE}$~=~4.0, significantly higher
than both the predictions of Big Bang nucleosynthesis (Cyburt et al. 2008)
and the observed lithium abundance in unevolved halo metal-poor stars
(Sbordone et al. 2010).
Dom\'{i}nguez et al (2004) and Kirby et al. (2012)
discovered 15 Li-rich, metal-poor giants in dwarf spheroidal galaxies near the 
Milky Way.
Kirby et al. provide a preliminary estimate of 0.85\% for the frequency of
super-litihum-rich red giants in dwarf spheroidal galaxies, consistent
with the 1\% frequency of occurance of lithium rich giants in Galactic
stellar populations (Brown et al. 1989, Ruchti et al. 2011).

Many authors have reprised discussions of lithium dilution and depletion as
stars evolved from the main sequence to the red giant branch and beyond
(see, for example, Kirby et al. 2012 and 2016, or D'Orazi et al. 2015 and references
therein). High lithium abundances in dwarf and giant stars are usually 
inconsistent with the standard model of stellar evolution, and 
several authors have proposed explanations for their origin. 
Monaco et al. (2012) offer two alternatives to explain the presence of 
a lithium-rich dwarf star in the globular cluster M4, either that the
star has preserved lithium from its formation (while other stars of the
same temperature in the cluster have not) or the excess lithium results
from mass transfer from a former asymptotic giant branch (AGB) companion.
Koch et al. (2011) consider that the lithium-rich turnoff star in NGC~6397
could have resulted from mass transfer from a former red giant companion star
that underwent cool-bottom processing (CPB; Wasserburg et al. 1995), 
but also propose other mechanisms, including possible diffusion 
of lithium to the surface. 

Wasserburg et al. (1995) and Boothroyd et al. (1995) computed models with
enhanced mixing to explore CPB and hot-bottom burning (HBB) in low mass stars.
They explored and predicted changes in the surface compositions of stars
due to deeper mixing in each case.
Sackmann \& Boothroyd (1999) examined the effect of CBP
on the abundances of light elements, and specifically on the abundance
of lithium in low mass red giants.  
They found that CBP in red giants could produce 
surface lithium abundances as high as A(Li)~=~4\footnote
{A(x) = 12 + log n(x)/n(H), where n(x) is the number density of atoms of species x
and n(H) is the number density of H atoms.}
 through the Cameron-Fowler
conveyor (Cameron \& Fowler 1971) with appropriate tuning of the mixing
mechanism.  

Palacios et al. (2001) proposed a scenario to explain anomalously high
lithium abundances in giants above the red giant luminosity function bump (LFB).  
Under certain conditions, newly synthesized lithium from the star's interior
adds to a star's energy production. 
Palacios et al.  argue that rotation-induced mixing can occur 
when the hydrogen burning zone
reaches the molecular weight discontinuity following the first dredge-up.
The extra mixing brings $^{7}$Be upwards, where it decays to $^{7}$Li.
The $^{7}$Li reacts with protons in a thin ''lithium burning shell,''
increasing the luminosity of the star by perhaps a third.
Palacios et al. suggest that the energy release from the 
lithium burning shell also enhances transport of processed material into the
convective zone, raising the surface lithium abundance.
The high surface lithium abundance is only temporary, however, since the star 
returns to its normal structure after a lithium burning episode, and the fresh 
lithium is eventually destroyed as it passes through the deep regions 
of the convective zone.

D'Orazi et al. (2015) identified a lithium-rich giant close to the 
LFB in the globular cluster NGC 362, but cannot
confirm whether the star is on the red giant branch or on the pre-zero-age
horizontal branch.
If the former, D'Orazi et al. suggest that the lithium could have been 
brought to the surface through enhanced mixing at the LFB bump, 
while if the latter, synthesis during the helium flash could have occurred.
Ruchti et al. (2011) found eight lithium-rich metal poor giants in
a survey of over 700 metal-poor giants, and determined that the
composition of the lithium-rich stars is similar to that of normal stars.
They suggest that the high lithium results either from extra mixing at the
LFB or from cool bottom processing in the AGB phase. 
Kirby et al. (2012) argue that lithium enrichment through enhanced mixing
following the luminosity function bump is a short-lived phase shared by all
metal-poor red giants.  They suggest that a high surface lithium phase 
that lasts 1\% of a giant's lifetime above the bump could explain the
fraction of high lithium stars observed.

More recently, Kirby et al. (2016) reported on a survey of nearly 1800 RGB and AGB
stars in 25 clusters.  They provide clear criteria for designating a star as 
lithium-rich and found that only one in 500 cluster RGB stars are lithium-rich,
while 1 in 60 AGB stars can be classified as lithium-rich. 
They reviewed scenarios to produce the lithium enhancement, including 
planet engulfment, self-enrichment through the Cameron-Fowler conveyor,
and binary mass-transfer.
Kirby et al. conclude that deep mixing at the helium core flash could explain
the high lithium abundances in HB and AGB stars, and that binary mass transfer
from post-core-flash stars could account for lithium-rich RGB stars.
  
We report here the detection of a prominent Li~I~6707 \AA~feature 
in the low mass, metal-poor, red giant star vZ~1050 in the globular cluster M3.
In Section 2 we describe our observations and data reduction, while in
Section 3 we discuss our analysis methods and the resulting abundances of 
elements, including lithium, in vZ~1050.  
Scenarios for the origin of the high lithium abundance are
reviewed in Section 4, and our conclusions are summarized in Section 5.

\section{Observations and Data Reductions}

The M3 giant vZ~1050 (J2000: 13h 43m 15.17s, +28$^{\circ}$ 21' 10.61")
in the globular cluster M3 (von Zeipel 1908; also known as K666 from Kustner 
1922 and SK 291 from Sandage \& Katem 1982) was included in 
a sample of stars observed with the Hydra multi-fiber spectrograph on the WIYN
3.5 m telescope on Kitt Peak.  
These observations were described and analyzed by Johnson et al. (2005), 
although the star vZ~1050 was not included in their analysis,
most likely because infrared photometry is not available for the star.  
The Hydra spectrum of vZ~1050 reveals a strong feature of Li I at 6707 \AA\ . 
Followup observations of vZ~1050 were obtained with the 3.5 m telescope at the
Apache Point Observatory in May, 2014, using the ARCES echelle
spectrograph (Wang et al. 2003).  These higher resolution, higher S/N ratio
observations are analyzed here.

The ARCES echelle spectrograph provides wavelength coverage from 
3500 \AA\ to 10,000 \AA\
with a resolving power of R~=~$\sim$31,500 (2.5 pixels) using the 1.6"~x~3.2" slit.
Two 3600s observations of vZ~1050 were obtained on 18 and 19  May 2014, for 
a total integration time of 7200 seconds.
Images were bias corrected, flat-fielded, cleaned and
extracted to one dimensional spectra,
wavelength calibrated, continuum normalized, and combined to a single
spectrum using IRAF\footnote{IRAF is distributed
by the National Optical Astronomy Observatory, which is operated by the
Association of Universities for Research in Astronomy, Inc., under
cooperative agreement with the National Science Foundation.}.
At wavelengths longer than 5000 \AA\ the signal-to-noise ratio of the final,
reduced spectrum is typically 75 near the center of the orders, but declines
to near 50 at the edges and to shorter wavelengths. 
The spectrum of vZ~1050 at the Li~I feature is shown in Figure 1, along
with synthetic spectra that will be discussed in the next section.

Tucholke et al. (1994) indicated that vZ~1050 is a 98\% probable member of M3,
and Pilachowski et al. (2000) reported a radial velocity of --150 km s$^{-1}$, 
consistent with the star's membership in M3.
Photometry from Sandage and Katem (1982) provides a V magnitude for the star
of V=14.23~$\pm$~0.031, with a B-V color of B-V=0.93~$\pm$~0.039.  
Tucholke et al. (1994) give a B magnitude of 15.16, consistent with the 
Sandage and Katem photometry. 
No K magnitude is available in SIMBAD.
We adopt a distance modulus (m-M)$_{V}$~=~15.07 and a reddening
E(B-V)~=~0.01 from Harris (1996, updated 2010).

\section{Analysis and Abundances}

We determine the abundance of several elements in vZ~1050 under the assumption
of local thermodynamic equilibrium using the MOOG spectrum synthesis program
(Sneden 1973, 2014 version).  We adopted the Yong et al. (2014) line list and gf values
for the analysis. Equivalent widths were measured using the splot routine
in IRAF and  equivalent widths were used to determine the abundances of
most species (exceptions are noted below). The equivalent widths for the 
spectral lines used are provided in Table~1, which also includes the gf values,
line excitation potentials, and the wavelengths of the spectral lines used.

Model atmospheres for vZ~1050 were interpolated in the MARCS\footnote
{The interpolation of models utilized code provided by
Masseron (2006, http://marcs.astro.uu.se/software.php).} 
grid (Gustafsson et al. 2008). 
Approximate initial atmospheric parameters were estimated from vZ~1050's 
color and V magnitude and were refined by minimizing trends 
of abundance vs. excitation potential using Fe I lines and vs. equivalent 
width using Fe I lines.  Fe I and II lines were used to fix the surface gravity. 
Ti I and Ti II lines could provide second estimate of the gravity,
but in practice, the Ti I lines scatter widely and the number of
reliable lines is small.  The resulting Ti abundances from Ti I
and Ti II lines are within a standard deviation.
Changes of 100K in temperature
and 0.2 km s$^{-1}$ in microturbulence from the adopted values 
are sufficient to introduce noticeable trends.

The final adopted atmospheric parameters for vZ~1050 are 
T$_{eff}$~=~4800~$\pm$~100~K,
log~g~=~1.8~$\pm$~0.3, [Fe/H]~=~-1.5~$\pm$~0.1, and
v$_{T}$~=~1.1~$\pm$~0.2~km~s$^{-1}$, consistent with
the star's position in M3's color magnitude diagram. 
Our derived abundances for vZ~1050 are given in Table 2, along with the
statistical standard deviation of the measurements for each speccies and the
number of lines used for each species.
For species with two or fewer lines measured, we have estimated the uncertainty
based on the dispersion for similar species.

\subsection{Metallicity}

The iron abundance of vZ~1050 was found to be [Fe/H]~=~--1.51~$\pm$~0.05
(standard error of the mean)
from 71~Fe~I lines.  The abundance from 16 Fe~II lines is similar:
[Fe/H]~=~--1.48~$\pm$~0.05.
Sneden et al. (2004) found an abundance of [Fe/H]~=~--1.58~$\pm$~0.06 from the 
Fe~I lines 
and [Fe/H]~=~--1.45~$\pm$~0.03 from Fe~II lines for their sample of 23 red giants
observed with the Keck telescope.
Our Fe~I and Fe~II abundances are consistent with those reported by 
Sneden et al., where determinations for individual stars range 
from --1.71$<$[Fe/H]$<$--1.50 for Fe~I
and from --1.50$<$[Fe/H]$<$--1.40 for Fe~II. 

Other metals analyzed include alpha elements Mg, Si, Ca, and Ti; the odd
element Sc; and the transition metals V, Cr, Mn, Ni, and Zn.  
The alpha element abundances are similar to other M3 giants reported by
Sneden et al. (2004), with [metal/Fe] ratios of [Ti I/Fe]~=~0.37, 
[Ti II/Fe]~=~0.22, [Ca/Fe]~=~0.23, [Mg/Fe]~=~0.11, and [Si/Fe]~=~0.25. 
As in the Sneden sample, Sc is slightly deficient.
The iron group metals scatter around [metal/Fe]~=~0.0.  In Figure 2, our derived [metal/Fe] 
ratios (filled circles with error bars) are plotted against atomic number along with the average
[metal/Fe] values for the sample of stars studied by Sneden et al. (2004) (open circles). 
As shown in Figure 2, we derive similar abundances for most elements, 
within the range of our uncertainties.

\subsection{Lithium}
The lithium abundance can be determined from the Li I 670 \AA\ doublet 
using spectrum synthesis. 
The strength of the Li I feature places vZ~1050 among the small 
population of Li-rich giants found in globular clusters.
In the region of the Li feature at 6700 \AA\  in the observed spectrum,
the S/N ratio is 70. 
We adopted the Reddy et al (2002) line list for the Li region;  this list
includes several known features that may blend with the Li~I feature,
including CN, Sm~II, V~I, Cr~I, Ce~II, Ti~I, and Ca~I.
The equivalent width of the Li~I feature is 95~m\AA\ , significantly stronger
than any of the possible blends, which make little contribution to the overall
strength of the feature.
Our best fit synthesized spectrum in Figure 2 indicates an LTE lithium abundance of 
A(Li)~=~1.52~$\pm$~0.05.
A second Li~I feature at 6103~\AA\ may be present in our spectrum, but 
at a temperature of T$_{eff}$~=~4800~K the 6103 \AA\ feature too weak to measure.
An upper limit of A(Li)~=~1.5 is consistent with the observed noise in this
spectral region.
A $^{6}$Li/$^{7}$Li ratio of 10\% was assumed in fitting the Li~I feature,
and this abundance represents the total lithium abundance in vZ~1050.
The resolution of our spectrum is not sufficient to measure the isotope ratio
directly.
Figure 2 also includes two additional synthetic spectra at values of 
A(Li)$_{LTE}$~=~1.4 (blue dashed) and 1.6 (green dashed).
Changes in the input lithium abundance of the synthetic spectrum of more than 
0.05~dex from our nominal value are clearly incompatible with the 
observed spectrum.

Corrections to the abundance of lithium for departures from local 
thermodynamic equilibrium (LTE) have been computed by Lind et al. (2009).
For stars with atmospheric parameters similar to those of vZ~1050, the
corrections are 0.08 dex, in the sense 
$\Delta$A(Li) = A(Li)$_{NLTE}$-A(Li)$_{LTE}$. 
Applying this correction to our LTE abundance gives an abundance
A(Li)$_{NLTE}$=1.60$\pm$0.17, taking into account the uncertainties in the
model parameters as well as the observational uncertainties.

The Kirby et al. (2016) criterion identifies stars as lithium-rich if their abundances 
are above A(Li)$_{NLTE}$~=~1.1 at the absolute magnitude of vZ~1050.  
The relatively high abundance confirms that vZ~1050 is a second 
Li-rich giant in M3, in addition to the RGB star IV-101 (Kraft et al. 1999).

\subsection{Oxygen}

The oxygen abundance can be determined from the [O I] feature at 6300 \AA\
using spectrum synthesis.  In this region, the S/N ratio of our spectrum
is 75.  The region is sprinkled with telluric 
O$_{2}$ and H$_{2}$O features, but the radial velocity of M3 fortuitously 
shifts the [O I] feature to a clean region between telluric lines.
Our stellar line list, derived from the Kurucz \& Bell (1995) line list 
via Sneden's ``fixlines.f'' program,
includes Ni I feature at $\lambda$6300.33 \AA\ and the nearby Sc II feature 
at $\lambda$6300.68 \AA\ , as well as features of Fe, Mn, Ce, and Cr 
that are too weak to be seen in the synthetic spectrum.

Comparing the observed spectrum with synthetic spectra computed for a range
of oxygen abundances, we derive a value of A(O)~=~7.7~$\pm$~0.1 
from our LTE, 1D analysis. 
This analysis was carried out using values of the carbon abundance 
[C/Fe]~=~--0.5, 0.0, and +0.5, with no change to the derived
abundance of oxygen.
The relatively high abundance of oxygen suggests 
that vZ~1050 is an oxygen-rich, first generation star.

\subsection{Sodium}

The sodium abundance serves as a proxy for determining whether vZ~1050 belongs
to the first generation of stars formed in M3 or to a subsequent generation.
Gratton et al. (2012) have shown that a weak sodium abundance is 
characteristic of first generation stars, while subsequent generations form
with higher sodium abundances.

The sodium abundance in vZ~1050 has been determined from three spectral lines,
5688, 8183, and 8194 \AA\ . The lines at 6154 and 6160 \AA\ are too weak 
to measure at the S/N ratio of our spectrum, and the 5682 feature is 
unmeasurable due to a defect in the spectrum.  
Atomic line parameters for the 8183 and 8194 \AA\ features, which are not
included in the Yong et al. line list,  were obtained from
Dobrovolskas et al. (2014).  
Non-LTE corrections from Dobrovolskas et al. and Lind et al. (2011)
were applied to the LTE abundances.
The 8183 and 8194 \AA\ features are particularly sensitive to NLTE effects,
with corrections of --0.45 dex, while the 5688 \AA\ feature is
relatively less sensitive, with an NLTE correction of --0.15 dex.
After the NLTE corrections are applied, the three features used are in
close agreement and suggest the abundance of sodium is
[Na/Fe] = --0.5~$\pm$0.1, comparable to the lowest sodium abundances found by 
Johnson et al. (2005) for M3 giants\footnote{Johnson et al. employed NLTE 
corrections from Gratton et al. (2000).  For comparison of vZ~1050 with 
other M3 giants, we have used the Johnson et al. LTE sodium abundances 
with the NLTE corrections from Lind et al. 2011).}.  

The low abundance of sodium in vZ~1050 confirms that it belongs to the
first generation of stars formed in M3. 
In contrast, the lower oxygen abundance and higher sodium abundance of
M3's other Li-rich giant IV-101 (Kraft et al. 1999) suggest it belongs
to a subsequent generation.
Both stars are plotted with other M3 giants from Sneden et al. (2004) in Figure 3,
showing [Na/Fe] vs. [O/Fe].  Note that to be consistent with published work, 
we have plotted the LTE abundance for [Na/Fe] without NLTE corrections.
The two stars clearly arise from different populations in M3.

\subsection{Heavy Elements}

While our spectrum extends below $\lambda$4000 \AA, we have (mostly) 
restricted our line list to transitions above $\lambda$5000 \AA\ to avoid
regions with low S/N ratio and heavy line blending.
Specific features used for the determination of heavy element abundances
are included in Table 1, and the derived abundances from an LTE equivalent 
width analysis are included in Table 2, along with the number of lines used.

Sneden et al. (2004) include determinations of Ba, La, and Eu
abundances for 23 giants in M3.  They find average abundances of
[Ba/Fe]~=~+0.21, [La/Fe]~=~+0.09, and [Eu/Fe]~=~+0.54, with star-to-star
variations of several tenths of a dex.  Within uncertainties, the Ba, La,
and Eu abundances we obtain are consistent with the spread of abundances
observed by Sneden et al.~(2004). On average, they found the barium and lanthanum
abundances in
M3 giants to be enhanced compared to field stars with the same metallicity.

In addition to Ba, La, and Eu, we have determined abundances for
Sr (from Sr~I $\lambda$4607~\AA), Y II, Ce II, Nd II, 
and Sm II. Ce, Nd, and Sm are all high, consistent with the abundances 
of Ba and La.  The abundances of Sr and Y appear to be low, although 
the number of lines for these species is small.  
Overall, the neutron capture element abundances for
vZ~1050 appear to fall within the normal range for M3 giants, and do not 
show a notable excess compared to other M3 giants.
The normality of these elements hints that mass
transfer from a former AGB companion undergoing
thermal flashes is unlikely to be the source of the 
enhanced lithium.

\subsection{Uncertainties}
The sensitivity to
errors in the atmospheric parameters listed in Table 3 was determined by 
varying each parameter by the amount we estimate to be the uncertainty 
in that parameter, and recalculating the abundance.
The lithium abundance A(Li) is relatively sensitive to temperature,
varying by 0.14 dex per 100 K change in temperature.  Our results are
relatively insensitive to errors in [Fe/H] and microturbulence.  The abundances 
of Fe~II and [O~I] are modestly sensitive to the adopted surface gravity.

\section{Results}

The star vz~1050 in M3 joins a relatively small group of lithium-rich 
giants known in globular clusters, including its sibling IV-101 in the same
cluster (Kraft et al. 1999).
In Figure 4, we plot the location of vZ~1050 in the M$_{V}$,(B-V)$_{0}$
color-magnitude diagram of M3, using photometric data for M3 giants from 
Rey et al. (2001).  
Since the Rey et al. fields do not include vZ~1050, its V magnitude and B-V color
are taken from Sandage \& Katem (1982).  
These authors indicate that the uncertainty
on the B-V color of  vZ~1050 is 0.039 mag, higher than its separation from the
ridgeline of RGB stars.
The absolute magnitudes and dereddened colors have been obtained using
a distance modulus of m-M = 15.07 and reddening E(B-V) = 0.01 from
Harris (1996, updated in 2010).
The position of vZ~1050 relative to the red giant branch suggests the star
is likely to be an asymptotic giant branch (AGB) star.

Also plotted in Figure 4 are the positions of the several previously known 
lithium-rich giants in globular clusters.
Photometry for M3 IV-101 (Kraft et al. 1999) has been corrected using
the same distance modulus and reddening as for vZ~1050.
Photometry for NGC 362 giants 15370 (D'Orazi et al. 2015) and 
V2 (Frogel et al. 1983) has been corrected using a distance modulus 
m-M = 14.82 and reddening E(B-V) = 0.05 from the 
Harris Catalog (1996, 2010 edition).
We note that Eggen (1972) found an amplitude of $\Delta$V = 0.8 mag 
and a range of $\Delta$(B-V) = 0.1 with a period of 90 days for this 
red variable star.
For A-96 in NGC 4590 (Ruchti et al. 2011), we used a distance modulus
m-M = 15.21 and E(B-V) = 0.05, again from the Harris Catalog, to
plot the photometry of Alcaino (1977). 
Stars identified by Kirby et al. (2016) are also included.

For guidance, we have also plotted isochrones for metallicities of
[Fe/H] = -2.2 (blue), [Fe/H]=--1.7 (black), and [Fe/H]--1.3 (red) from 
Bertelli et al. (2008).
The isochrones are computed at Y=0.26 and with log (Age) = 10.15.
For reference, Harris (1996, updated 2010) lists the metallicity of NGC 4590 to be
[Fe/H]=--2.2, of NGC 362 to be [Fe/H]=--1.3, and of M3 to be [Fe/H]=--1.5.

Kirby et al. (2016) summarize the evolutionary state of the
known lithium-rich globular cluster giants.
Four lie on the lower RGB, while others may be upper RGB or AGB stars.
As D'Orazi et al. (2015) note, star 15370 in NGC~362, with a lithium abundance A(Li)~=~ 2.4, lies clearly on its cluster's red giant branch, just above the LFB.  
NGC 362's V2, NGC 4590's A-96,
and M3's IV-101 all lie near the tips of their respective red giant branches.
In contrast, vZ~1050, with A(Li)=1.5, lies blueward of M3's RGB, and can 
be classified as a likely early AGB star.  
Kirby et al. (2016) find that only 0.2\% of RGB stars are lithium-rich, while a much
higher percentage, 1.6\%, of AGB stars qualify as lithium-rich.

As discussed in Section 1, several pathways have been described in the literature to 
account for the relatively rare, lithium-rich giants in globular clusters. 
Kirby et al. (2012) argue that since 1\% of cluster stars more luminous 
than the red giant LFB display Li abundances higher than predicted by 
stellar evolution models,
all such red giants may evolve through a short-lived phase resulting from
in situ lithium production, followed by rapid lithium depletion.
Kirby et al. (2016) provide a more nuanced analysis, suggesting a scenario
in which the higher fraction of lithium-rich AGB stars results from a common 
but short-lived enrichment phase associated with or following the helium flash. 
While the Cameron-Fowler conveyor provides a transport mechanism,
the proximate cause of the required deep mixing remain vague.
Kirby et al. suggest that the rarity of lithium-rich stars at earlier evolutionary phases can be explained 
if they result from mass transfer from more 
massive, and more evolved companions.  
The absence of any s-process enhancement in M3~ vZ~1050, however, may not be consistent with this scenario.

\section{Summary}
In summary, the star vZ~1050 in the globular cluster M3 is found to be 
lithium-rich, with a lithium abundance of A(Li)$_{NLTE}$~=~1.6~$\pm$~0.05.
The star's radial velocity confirms its membership in the cluster, and its 
absolute magnitude and color suggest that vZ~1050 is ascending the
red giant branch for the second time as an AGB star.
The abundances of the alpha, transition metal, and neutron capture elements
in the atmosphere of  vZ 1050 are typical of the abundances of other M3 giants.
The star does show a high oxygen and low sodium abundance compared to 
other M3 giants, however, consistent with vZ~1050 belonging to the first generation
of stars formed in M3.
The origin of its lithium enhancement likely results
from the in situ operation of the Cameron-Fowler
mechanism. 

\acknowledgments

Based on observations obtained with the Apache Point Observatory 3.5-meter telescope, 
which is owned and operated by the Astrophysical Research Consortium.
We are particularly grateful to
to Jack Dembicky for his assistance during our observing run.
This research has made use of the NASA Astrophysics Data System 
Bibliographic Services, as well as the SIMBAD database, operated at CDS,
Strasbourg, France.
We thank Eric Ost for implementing the model atmosphere 
interpolation code.
C.A.P. acknowledges the generosity of the Kirkwood Research Fund at 
Indiana University.
Finally, we are grateful to the referee, whose comments have helped to improve this paper.

\vspace{5mm}


\clearpage

\bibliographystyle{plainnat}

\clearpage

\begin{figure}
\epsscale{.80}
\plotone{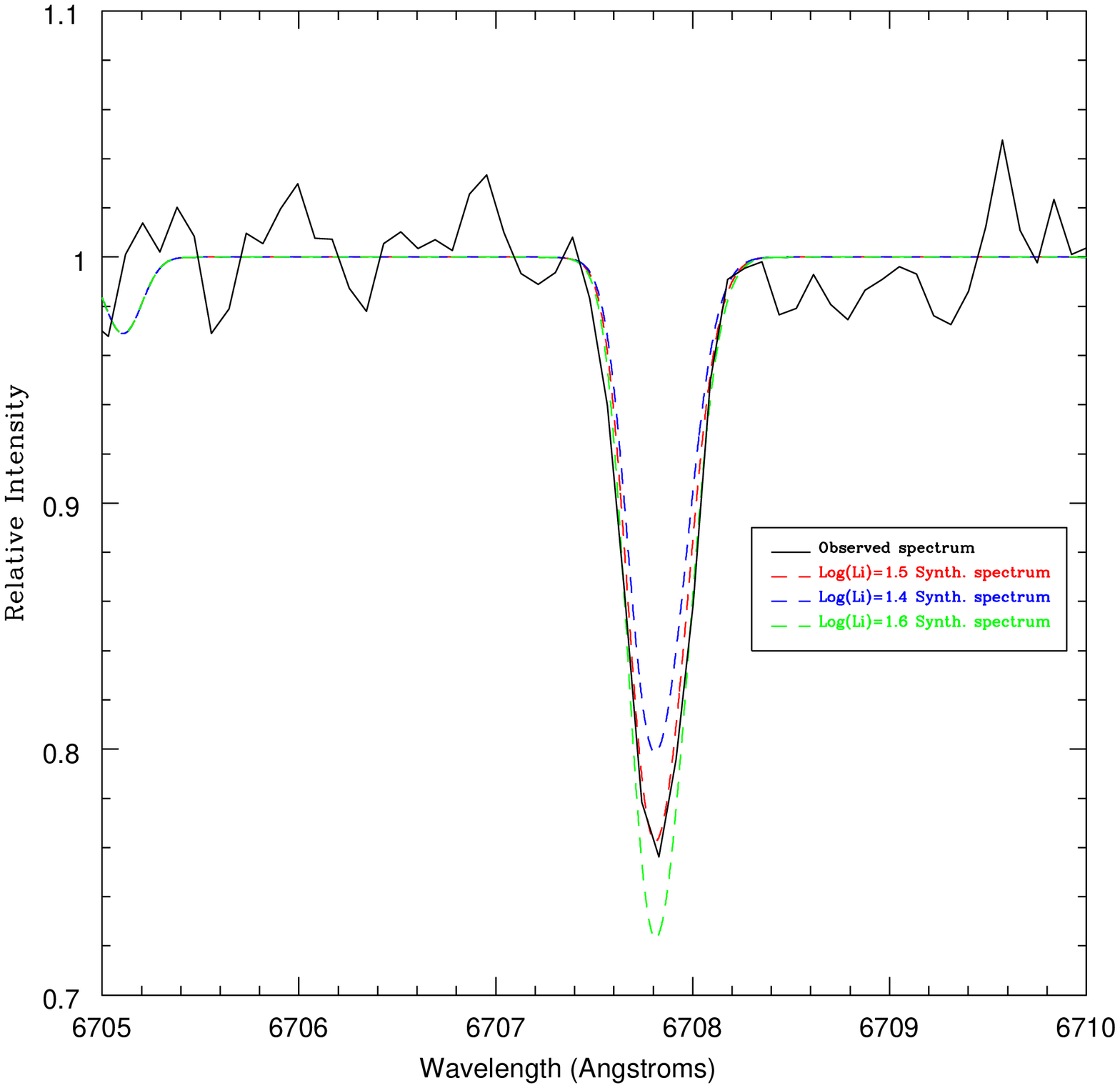}
\caption{Observed and synthetic spectra of the Li~I feature in vZ~1050.  
The observed spectrum is shown with a solid line and synthesized lithium spectra at A(Li) = 1.4, 1.5, and 1.6 (in LTE) are shown as dashed blue, red, and green lines, respectively.  
The derived LTE lithium abundance is A(Li)$_{LTE}$~=~1.5~$\pm$~0.05 at 
T$_{eff}$~=~4800~$\pm$~100~K.  Including the NLTE correction from Lind et al.
(2009), we obtain A(Li)$_{NLTE}$~=~1.6. }
\end{figure}

\begin{figure}
\epsscale{.80}
\plotone{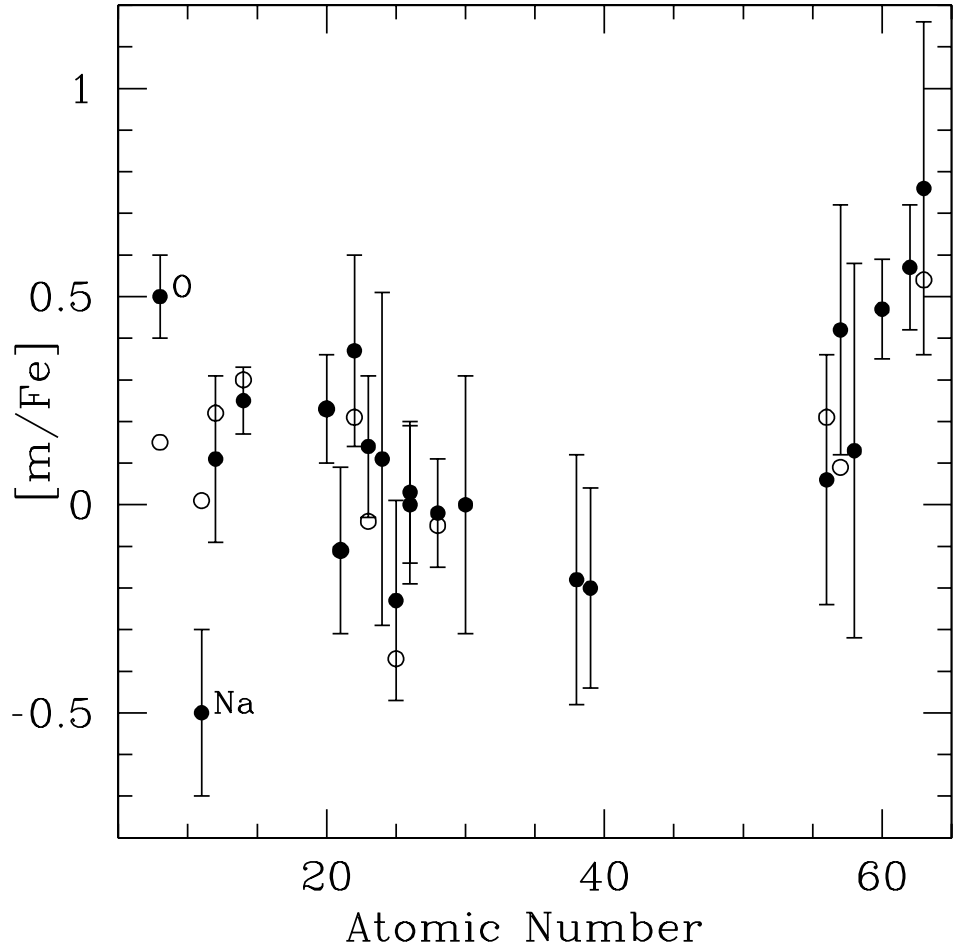}
\caption{Derived abundances [m/Fe] for the M3 AGB stars vZ~1050 vs. atomic number (filled symbols).  The average [m/Fe] values for M3 giants from Sneden et al. 
(2004) are plotted with open symbols for comparison.  Note that oxygen is significantly
more abundant and sodium is significantly less abundant than the average for M3.
Other species are consistent with the Sneden et al. averages for M3 within our
uncertainties.}
\end{figure}

\begin{figure}
	\epsscale{.80}
	\plotone{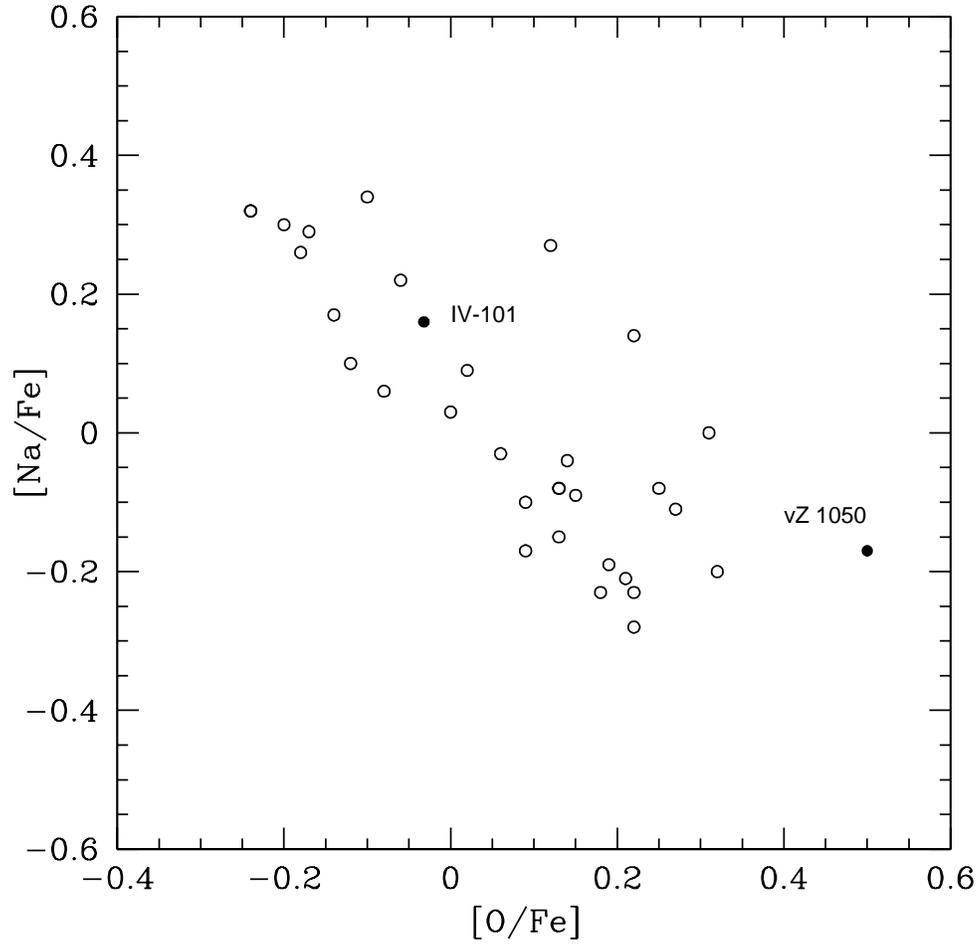}
	\caption{[Na/Fe] vs. [O/Fe] for vZ~1050, IV-101, and additional red giants in M3 from Sneden et al. (2004).
	Both lithium rich giants are plotted as filled circles and labeled.
	The [Na/Fe] abundances plotted are LTE to avoid
	confusion due to different NLTE corrections.}
\end{figure}

\begin{figure}
\epsscale{.80}
\plotone{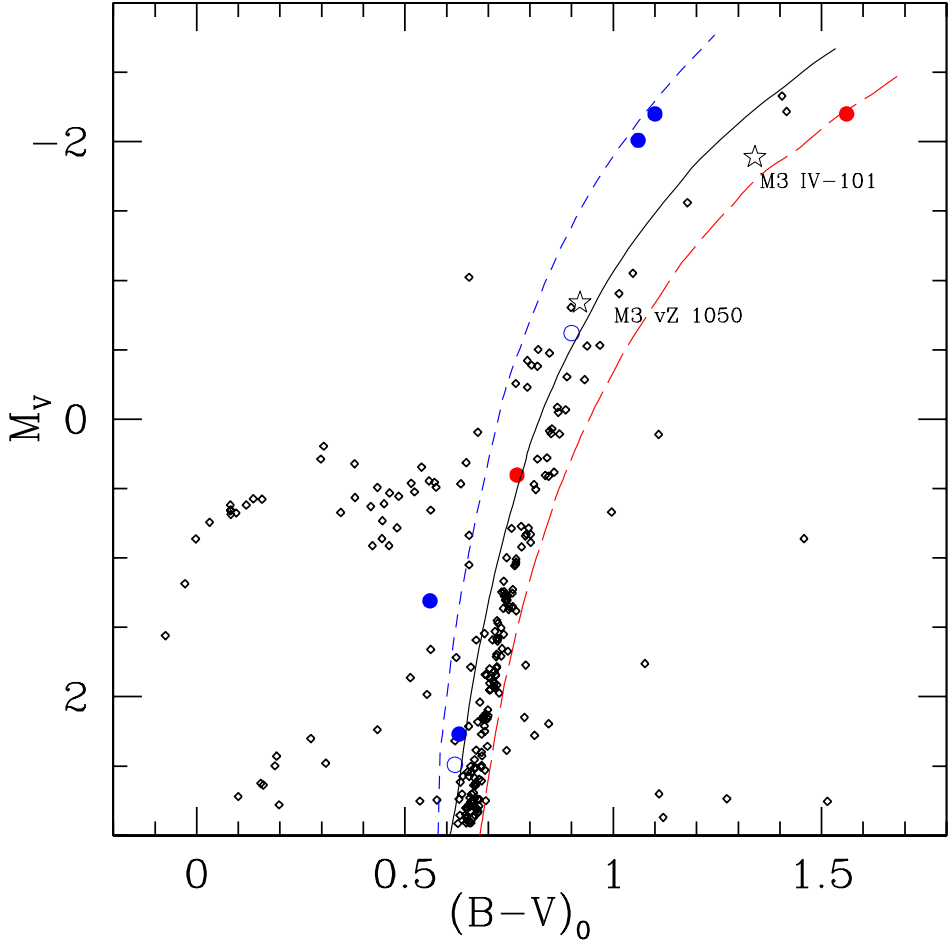}
\caption{The M$_{V}$ vs. (B-V)$_{0}$ color-magnitue diagram of M3 using 
photometry from Rey et al. (2001).  Isochrones from Bertelli et al. (2008) for [Fe/H]~=~--2.0
(blue, small-dashed line), [Fe/H]~=~--1.7 (solid black line), and [Fe/H]~=~--1.3
(red, large-dashed line) are shown to guide the eye. 
The two lithium-rich stars in M3 are shown with open starred symbols and identified by name.  Other known Li-rich stars are shown with blue symbols for more metal-poor clusters (M68, NGC 5053, NGC 5897, and M30, Kirby et al. 2016; and NGC 4590, Ruchti et al. 2011) and with red symbols for the more
metal-rich cluster NGC 362 (Frogel et al. 1983 and D'Orazi et al. 2015).  For some stars identified by Kirby et al., B-V colors are not available and have been estimated based
on the stars' temperatures; these cases are shown as blue open symbols.}
\end{figure}

\clearpage

\begin{deluxetable}{lcccc}
\tabletypesize{\scriptsize}
\tablecaption{Equivalent Widths of Spectral Features in vZ~1050\label{table_eqw}}
\tablewidth{0pt}
\tablehead{ & \colhead{Wavelength} & E.P. &  & \colhead{EQW} \\
\colhead{Species} & \colhead{(\AA\ )} & \colhead{(eV)} & \colhead{Log gf} &
\colhead{(m\AA\ )}
}
\startdata
Fe I  & 4839.55	& 3.27 & --1.84	& 27 \\
Fe I  & 4892.87	& 4.22 & --1.21	& 13 \\
Fe I  & 4917.24	& 4.19 & --1.27	& 32 \\
Fe I  & 4924.77	& 2.28 & --2.29	& 66 \\
Fe I  & 4961.92	& 3.64 & --2.34	& 12 \\
Fe I  & 5014.94	& 3.94 & --0.32	& 63 \\
Fe I  & 5044.21	& 2.85 & --2.03	& 49 \\
Fe I  & 5060.08	& 0.00 & --5.45	& 60 \\
Fe I  & 5217.39	& 3.21 & --1.18	& 59 \\
Fe I  & 5242.49	& 3.63 & --0.98	& 42 \\
Fe I  & 5243.78	& 4.26 & --1.01	& 24 \\
Fe I  & 5253.46	& 3.28 & --1.63	& 35 \\
Fe I  & 5288.53	& 3.69 & --1.53	& 21 \\
Fe I  & 5364.86	& 4.44 &    0.21 & 63 \\
Fe I  & 5367.48	& 4.41 &    0.43 & 54 \\
Fe I  & 5398.29	& 4.45 & --0.68	& 29 \\
Fe I  & 5522.45	& 4.21 & --1.45	& 10 \\
Fe I  & 5525.54	& 4.23 & --1.15	& 18 \\
Fe I  & 5560.22	& 4.44 & --1.07	& 24 \\
Fe I  & 5618.64	& 4.21 & --1.32	& 20 \\
Fe I  & 5662.51	& 4.18 & --0.59	& 52 \\
Fe I  & 5701.56	& 2.56 & --2.22	& 63 \\
Fe I  & 5717.84	& 4.29 & --0.94	& 32 \\
Fe I  & 5731.77	& 4.26 & --1.08	& 21 \\
Fe I  & 5753.12	& 4.26 & --0.71	& 34 \\
Fe I  & 5775.09	& 4.22 & --1.14	& 14 \\
Fe I  & 5806.73	& 4.61 & --0.91	& 12 \\
Fe I  & 5816.37	& 4.55 & --0.62	& 29 \\
Fe I  & 5852.23	& 4.55 & --1.17	& 16 \\ 
Fe I  & 5862.37	& 4.55 & --0.50 & 46 \\
Fe I  & 5916.25	& 2.45 & --2.99	& 31 \\
Fe I  & 5930.19	& 4.65 & --0.29	& 33 \\
Fe I  & 5934.67	& 3.93 & --1.15	& 50 \\
Fe I  & 5956.71	& 0.86 & --4.60	& 51 \\
Fe I  & 5976.79	& 3.94 & --1.33	& 36 \\
Fe I  & 6003.02	& 3.88 & --1.08	& 41 \\
Fe I  & 6007.97	& 4.65 & --0.82	& 20 \\
Fe I  & 6008.57	& 3.88 & --0.96	& 58 \\
Fe I  & 6027.06	& 4.08 & --1.23	& 28 \\
Fe I  & 6056.01	& 4.73 & --0.42	& 27 \\
Fe I  & 6079.02	& 4.65 & --0.95	& 28 \\
Fe I  & 6082.72	& 2.22 & --3.57	& 26 \\
Fe I  & 6089.57	& 4.58 & --1.28	& 21 \\
Fe I  & 6151.62	& 2.18 & --3.30	& 45 \\
Fe I  & 6157.73	& 4.08 & --1.29	& 28 \\
Fe I  & 6200.32	& 2.61 & --2.44	& 56 \\
Fe I  & 6229.23	& 2.84 & --2.85	& 19 \\
Fe I  & 6240.65	& 2.22 & --3.23	& 35 \\
Fe I  & 6246.33	& 3.60 & --0.73	& 55 \\
Fe I  & 6270.23	& 2.86 & --2.46	& 44 \\
Fe I  & 6301.51	& 3.65 & --0.72	& 77 \\
Fe I  & 6322.69	& 2.59 & --2.43	& 61 \\
Fe I  & 6336.82 & 3.68 & --0.92	& 65 \\
Fe I  & 6344.15	& 2.43 & --2.92	& 44 \\
Fe I  & 6355.03 & 2.84 & --2.40	& 40 \\
Fe I  & 6380.75	& 4.19 & --1.37	& 22 \\
Fe I  & 6411.66	& 3.65 & --0.60 & 79 \\
Fe I  & 6498.94	& 0.96 & --4.70 & 70 \\
Fe I  & 6518.37	& 2.83 & --2.46	& 42 \\
Fe I  & 6575.02	& 2.59 & --2.73	& 48 \\
Fe I  & 6593.88	& 2.43 & --2.42	& 76 \\
Fe I  & 6608.04	& 2.28 & --3.96	& 16 \\
Fe I  & 6609.12	& 2.56 & --2.69	& 46 \\
Fe I  & 6703.58	& 2.76 & --3.01	& 27 \\
Fe I  & 6726.67	& 4.61 & --1.10 & 21 \\
Fe I  & 6750.16	& 2.42 & --2.62	& 68 \\
Fe I  & 6806.86	& 2.73 & --3.14	& 29 \\
Fe I  & 6839.83	& 2.56 & --3.35	& 24 \\
Fe I  & 6843.66	& 4.55 & --0.85	& 20 \\
Fe I  & 7223.66	& 3.02 & --2.27	& 41 \\
Fe II & 4993.35	& 2.81 & --3.56	& 21 \\
Fe II & 5132.67	& 2.81 & --3.95	& 17 \\
Fe II & 5264.81	& 3.34 & --3.21	& 40 \\
Fe II & 5325.56	& 3.22 & --3.18	& 23 \\
Fe II & 5414.08	& 3.22 & --3.61	& 12 \\
Fe II & 5425.26	& 3.20 & --3.27	& 18 \\
Fe II & 5534.85	& 3.25 & --2.75	& 44 \\
Fe II & 5991.38	& 3.15 & --3.56	& 16 \\
Fe II & 6149.25	& 3.89 & --2.73	& 16 \\
Fe II & 6247.56	& 3.89 & --2.33	& 29 \\
Fe II & 6369.46	& 2.89 & --4.21	& 20 \\
Fe II & 6416.93	& 3.89 & --2.70 & 19 \\
Fe II & 6432.68 & 2.89 & --3.58	& 32 \\
Fe II & 6456.39	& 3.90 & --2.10 & 42 \\
Fe II & 6516.08	& 2.89 & --3.38	& 45 \\
Fe II & 7711.72	& 3.90 & --2.54 & 31 \\
Na I & 5688.22	& 2.10 & --0.37 & 31 \\
Na I & 8183.26 & 2.10 &   0.23  & 77 \\
Na I & 8194.82 & 2.10 &   0.49  & 100 \\
Mg I & 5711.09	& 4.34 & --1.73 & 49 \\
Si I & 5690.43 & 4.93 & --1.87 & 20 \\
Si I & 5772.15 & 5.08 & --1.75 & 11 \\
Si I & 5948.55 & 5.08 & --1.23 & 35 \\
Si I & 6155.14 & 5.62 & --0.86 & 32 \\
Si I & 7415.96 & 5.62 & --0.71 & 39 \\
Ca I & 5260.39	& 2.52 & --1.72	& 17 \\
Ca I & 5261.71	& 2.52 & --0.58	& 58 \\
Ca I & 5512.99 & 2.93 & --0.45	& 25 \\
Ca I & 5581.98	& 2.52 & --0.56	& 61 \\
Ca I & 5590.13	& 2.51 & --0.57	& 53 \\
Ca I & 5601.29	& 2.52 & --0.52	& 52 \\
Ca I & 6161.29	& 2.52 & --1.27	& 21 \\
Ca I & 6166.44	& 2.52 & --1.14	& 28 \\
Ca I & 6169.04	& 2.52 & --0.80 & 40 \\
Ca I & 6169.56	& 2.52 & --0.48	& 62 \\
Ca I & 6471.67	& 2.52 & --0.69	& 54 \\
Ca I & 6499.65	& 2.52 & --0.82	& 55 \\
Ca I & 6717.69	& 2.71 & --0.52	& 65 \\
Sc II & 5239.82	& 1.45 & --0.70	& 36 \\
Sc II & 5526.82	& 1.77 &     0.20 & 67 \\
Sc II & 5657.88	& 1.51 & --0.28	& 63 \\
Sc II & 5669.04	& 1.50 & --1.00	& 26 \\
Sc II & 6245.62	& 1.51 & --1.05	& 32 \\
Ti I & 4820.41 & 1.50 & --0.44 & 37 \\
Ti I & 5009.66 & 0.02 & --2.26 & 27 \\
Ti I & 5016.17 & 0.85 & --0.57 & 58 \\
Ti I & 5022.87 & 0.83 & --0.43 & 62 \\
Ti I & 5147.48 & 0.00 & --2.01 & 34 \\
Ti I & 5219.71 & 0.02 & --2.29 & 24 \\
Ti I & 5662.16 & 2.32 & --0.11 & 16 \\
Ti I & 5866.46 & 1.07 & --0.84 & 31 \\
Ti I & 5922.12 & 1.05 & --1.47 & 25 \\
Ti I & 5937.81 & 1.07 & --1.89 & 11 \\
Ti I & 5941.75 & 1.05 & --1.52 & 14 \\
Ti I & 5953.16 & 1.89 & --0.33 & 20 \\
Ti I & 5965.83 & 1.88 & --0.41 & 19 \\
Ti I & 5978.55 & 1.87 & --0.50 & 20 \\
Ti I & 6743.12 & 0.90 & --1.63 & 19 \\
Ti II & 4865.62 & 1.12 & --2.81 & 39 \\
Ti II & 5154.08 & 1.57 & --1.92 & 59 \\
Ti II & 5185.91 & 1.89 & --1.50 & 58 \\
Ti II & 5336.79 & 1.58 & --1.63 & 68 \\
Ti II & 5381.03 & 1.59 & --2.08 & 58 \\
Ti II & 5418.78 & 1.58 & --2.11 & 51 \\
V I & 5627.64 & 1.08 & --0.37 & 21 \\
V I & 5670.86 & 1.08 & --0.42 & 23 \\
V I & 5727.06 & 1.08 & --0.01 & 25 \\
V I & 6081.45 & 1.05 & --0.58 & 16 \\
V I & 6090.22 & 1.08 & --0.06 & 22 \\
V I & 6119.53 & 1.06 & --0.32 & 13 \\
V I & 6199.19 & 0.29 & --1.28 & 19 \\
V I & 6216.36 & 0.28 & --0.81 & 20 \\
V I & 6251.82 & 0.29 & --1.34 & 12 \\
Cr I & 4651.29 & 0.98 & --1.48 & 62 \\
Cr I & 4718.42 & 3.20 &    0.09 & 19 \\
Cr I & 4724.41 & 3.09 & --0.73 & 14 \\
Cr I & 4801.03 & 3.12 & --0.13 & 12 \\
Cr I & 4936.34 & 3.11 & --0.22 & 12 \\
Cr I & 5348.33 & 1.00 & --1.29 & 69 \\
Mn I & 4754.04 & 2.28 & --0.09 & 59 \\
Mn I & 5420.37 & 2.14 & --1.46 & 16 \\
Ni I & 4904.42 & 3.54 & --0.19 & 37 \\
Ni I & 4935.83 & 3.94 & --0.34 & 16 \\
Ni I & 5578.73 & 1.68 & --2.67 & 40 \\
Ni I & 5593.75 & 3.90 & --0.79 & 11 \\
Ni I & 5805.23 & 4.17 & --0.60 & 10 \\
Ni I & 5847.01 & 1.68 & --3.44 & 15 \\
Ni I & 6007.31 & 1.68 & --3.34 & 14 \\
Ni I & 6108.12 & 1.68 & --2.49 & 48 \\
Ni I & 6175.37 & 4.09 & --0.53 & 15 \\
Ni I & 6176.82 & 4.09 & --0.26 & 16 \\
Ni I & 6177.25 & 1.83 & --3.60 & 11 \\
Ni I & 6586.32 & 1.95 & --2.79 & 27 \\
Ni I & 6767.78 & 1.83 & --2.11 & 68 \\
Zn I & 4722.16 & 4.03 & --0.37 & 38 \\
Zn I & 4810.53 & 4.08 & --0.15 & 37 \\
Sr I & 4607.33 & 0.00 &   0.28 & 15 \\
Y II & 5087.42 & 1.08 & --0.16 & 40 \\
Y II & 5200.42 & 0.99 & --0.57 & 26 \\
Y II & 5509.91 & 0.99 & --1.01 & 46 \\
Ba II & 4554.03 & 0.00 &   0.14 & 107 \\
Ba II & 5853.69 & 0.60 & --0.91 & 73 \\
La II & 5114.56	& 0.23 & --1.03	& 16 \\
Ce II & 4486.91	 & 0.29 & --0.18 & 16 \\
Ce II & 4628.16	 & 0.52 &   0.14 & 18 \\
Ce II & 5187.46	 & 1.21 &   0.17 & 14 \\
Nd II & 4706.54	 & 0.00 & --0.71 & 32 \\
Nd II & 4825.48	 & 0.18 & --0.42 & 35 \\
Nd II & 4859.03 & 0.32 & --0.44	 & 33 \\
Nd II & 4959.12 & 0.06 & --0.80 & 29 \\
Nd II & 5212.36	 & 0.20 & --0.96 & 17 \\
Nd II & 5234.19	 & 0.55 & --0.51 & 27 \\
Nd II & 5249.58 & 0.97 &   0.20 & 22 \\
Nd II & 5293.16 & 0.82 &   0.10 & 35 \\
Nd II & 5319.81	 & 0.55 & --0.14 & 30 \\
Sm II & 4434.32 & 0.38 & --0.07 & 23 \\
Sm II & 4642.23 & 0.38 & --0.46 & 16 \\
Sm II & 4669.64 & 0.28 & --0.53 & 20 \\
Sm II & 4676.90 & 0.04 & --0.87 & 14 \\
Eu II & 6645.06	 & 1.38 &   0.12 & 25 
\enddata
\tablecomments{Table 1 is presented in its entirety in the electronic
edition of the Publications.  A portion is shown here for
guidance regarding its form and content.}
\end{deluxetable}
\clearpage

\begin{deluxetable}{lcccc}
\tabletypesize{\scriptsize}
\tablecaption{Abundances of the Elements in vZ~1050\label{table_abund}}
\tablewidth{0pt}
\tablehead{\colhead{Species} & \colhead{[m/H]} & \colhead{$\sigma$} 
& \colhead{N} & \colhead{[m/Fe]}
}
\startdata
Fe I        & --1.51 & 0.19 & 71 & 0.00  \\
Fe II       & --1.48 & 0.17 & 16 & 0.03  \\
O I         & --1.0  & 0.1  & 1  & 0.5   \\
Na I (LTE)  & --1.63 & 0.2  & 3  & --0.17 \\
Na I (NLTE) & --2.0  & 0.2  & 3  & --0.5 \\
Mg 1        & --1.40 & 0.2  & 1  & 0.11  \\
Si I        & --1.26 & 0.08 & 5  & 0.25  \\
Ca I        & --1.28 & 0.13 & 13 & 0.23  \\
Sc II       & --1.62 & 0.20 & 5  & --0.11 \\
Ti I        & --1.14 & 0.23 & 15 & 0.37   \\
Ti II       & --1.29 & 0.13 & 6  & 0.22  \\
V I         & --1.37 & 0.17 & 9  & 0.14  \\
Cr I        & --1.40 & 0.40 & 6  & 0.11  \\
Mn I        & --1.74 & 0.24 & 2  & --0.23 \\
Ni I        & --1.53 & 0.13 & 13 & --0.02 \\
Zn I        & --1.51 & 0.31 & 2  & 0.00  \\
Sr I        & --1.69 & 0.3  & 1  & --0.18 \\
Y II        & --1.71 & 0.24 & 3  & --0.20 \\
Ba II       & --1.45 & 0.3  & 2  & 0.06 \\
La II       & --1.09 & 0.3  & 1  & 0.42 \\
Ce II       & --1.38 & 0.45 & 3  & 0.13 \\
Nd II       & --1.04 & 0.12 & 9  & 0.47 \\
Sm II       & --0.94 & 0.15 & 4  & 0.57 \\
Eu II       & --0.75 & 0.4  & 1  & 0.76
\enddata
\end{deluxetable}  

\begin{deluxetable}{lcccccc}
\tabletypesize{\scriptsize}
\tablecaption{Sensitivities of Abundance Parameters for vZ~1050\label{table_sensitive}}
\tablewidth{0pt}
\tablehead{ & \colhead{T$_{eff}$$\pm$~100} & \colhead{log g$\pm$0.50} & 
\colhead{[M/H]$\pm$0.10} & \colhead{v$_{t}$$\pm$0.20} & & \\
\colhead{Species} & \colhead{(K)} & \colhead{(cgs)} & \colhead{(dex)} 
& \colhead{(km~s$^{-1}$)} & \colhead{Obs.} & \colhead{Combined}
}
\startdata
{[Fe/H]}$_{I}$ & 0.10 & $< 0.01$ & $< 0.01$ & 0.03 & 0.02 & 0.11 \\
{[Fe/H]}$_{II}$ & 0.05 & 0.12 & $< 0.01$ & 0.03 & 0.04 & 0.14     \\
A(Li) & 0.14 & $< 0.01$ & $< 0.01$ & $< 0.01$ & 0.1 & 0.17 \\
{[O/H]} & 0.14 & 0.11 & $< 0.01$ & 0.01 & 0.1 & 0.20 \\
{[Na/H]} & 0.07 & $< 0.01$ & $< 0.01$  & 0.01 & 0.2 & 0.21
\enddata
\end{deluxetable}
\clearpage

\end{document}